\documentclass[runningheads]{llncs}
\usepackage{array}
\newcolumntype{C}[1]{>{\centering\let\newline\\\arraybackslash\hspace{0pt}}m{#1}}
\usepackage{amsfonts}
\usepackage{amsmath, amssymb}
\usepackage{esint}
\usepackage{graphicx}
\begin{document}
\title{Connectivity-Driven Brain Parcellation via Consensus Clustering}
%
%
%
\author{Anvar Kurmukov\inst{1,2} \and Ayagoz Mussabayeva\inst{1} \and Yulia Denisova\inst{1} \and Daniel Moyer\inst{4} \and Boris Gutman\inst{3,1} }

\authorrunning{Anvar Kurmukov et al.}
%
\institute{
The Institute for Information Transmission Problems\and
National Research University Higher School of Economics\and
Illinois Institute of Technology\and
University of Southern California
}
%
\maketitle              
\begin{abstract}
We present two related methods for deriving connectivity-based brain atlases from individual connectomes. The proposed methods exploit a previously proposed dense connectivity representation, termed continuous connectivity, by first performing graph-based hierarchical clustering of individual brains, and subsequently aggregating the individual parcellations into a consensus parcellation. The search for consensus minimizes the sum of cluster membership distances, effectively estimating a pseudo-Karcher mean of individual parcellations. We assess the quality of our parcellations using (1) Kullback-Liebler and Jensen-Shannon divergence with respect to the dense connectome representation, (2) inter-hemispheric symmetry, and (3) performance of the simplified connectome in a biological sex classification task. We find that the parcellation based-atlas computed using a greedy search at a hierarchical depth 3 outperforms all other parcellation-based atlases as well as the standard Dessikan-Killiany anatomical atlas in all three assessments. 


\end{abstract}

\section{Introduction}

The ability to quantify how the human brain is interconnected in vivo has opened the door to a number of possible analyses. In nearly all of these, brain parcellation plays a crucial role. Variations in parcellation significantly impact connectome reproducibility, derived graph-theoretical measures, and the relevance of connectome measures with respect to biological questions of interest \cite{Petrov_35_connectomes}. A natural approach is then to use individual densely sampled connectomes to drive the parcellation directly, leading to a more compact, connectivity-aware set of brain regions and resulting graph, as done in e.g. \cite{multimodal_parcel}. A comprehensive review of parcellation methods and their effects on the derived connectome quality is given in \cite{Arslan_parcellation_compare}. Because individual connectivity data is at once very informative and highly redundant, there is a great flexibility in how parcels can be derived from dense, highly resolute graphs. It is possible for example to  derive (1) a unified population-based atlas, (2) individual-level parcellations with cross-subject label mapping, or (3) individual parcellations with no inter-subject label correspondence. While the first approach is appealing for its simplicity and ease of interpretation, the second and third may enable the researcher to reveal some individual aspect of the connectome that is lost in the aggregate atlas. 

In this work, we attempt to bridge these three approaches by first constructing maximally flexible hierarchical parcellations, and then finding a unifying set of labels and parcels to maximize individual agreement. We use the a continuous representation of a brain connectivity \cite{ConCon} as our initial dense connectome representation. Continuous connectivity is a parcellation-free representation of tractography-based, or ``structural'' connectomes that is based on 
the Poisson point process. Once individual parcellations are computed, we obtain a group-wise parcellation using partition ensemble algorithm. We access quality of the resulting parcellations in three ways. (1) We use the continuous connectome framework to  compare parcellation-approximate and exact edge distribution functions. (2) We compare perfomance of the resulting graphs on a gender classification task. (3) We also show that without any explicit knowledge of brain geometry and based solely on graph connectivity we obtain comparatively symmetric parcellations.

\section{Methods}

\subsection{Continuous Connectome}
The continuous connectome model (ConCon) treats each tract 
as an observation of an inhomogeneous symmetric Poisson point process with the intensity function given by
\begin{equation}
\lambda: \Omega \times \Omega \rightarrow \mathbb{R}^+,
\end{equation}
where $\Omega$ denote union of two disjoint toplogically spherical brain hemispheres, representing cortical white matter boundaries. In practice, ConCon uses cortical mesh vertices as nodes of connectivity graph. 
From such a representation, a ``discrete'' connectivity graph could be computed from any particular cortical parcellation $P$.
We follow definitions from \cite{ConCon} and call $P = \{E_i\}_{i=1}^{N}$ a parcellation of $\Omega$ if 
$E_1\ldots E_k \subseteq \Omega$ such that $ \cup_i E_i = \Omega$, and $N$ is the number of parcels (ROIs). Edges between regions $E_i$ and $E_j$ can then be computed by integration of the intensity function:
\begin{equation}
\mathcal{C}(E_i, E_j) = \iint_{E_i, E_j} \lambda(x,y) dxdy,
\end{equation}
Due to properties of the Poisson Process, $\mathcal{C}(E_i,E_j)$ is
the expectation of the number of observed tracts between $E_i$ and
$E_j$. In the context of connectomics, this is the expected edge
strength.

 




\subsection{Graph Clustering} Once we obtain all individual continuous connectomes, we partition each independently into a set of disjoint communities. For graph clustering we use the Louvain modularity algorithm \cite{louvain}, as it has shown good results in multiple neuroimaging studies \cite{neuro_louvain1}, \cite{neuro_louvain2}, \cite{neuro_clustering}, \cite{modular_brain}. This algorithm consist of two steps. The first step combines locally connected nodes into communities, while the second step builds new meta graph. The nodes of the meta-graph are communities from the previous step, and the edges are defined as the sum of all inter-community connections of the new nodes. The algorithm in \cite{louvain} cycles over these steps iteratively, converging when further node clustering leads to no increase in modularity.
\begin{figure}[!h]
\center{\includegraphics[width=6cm]{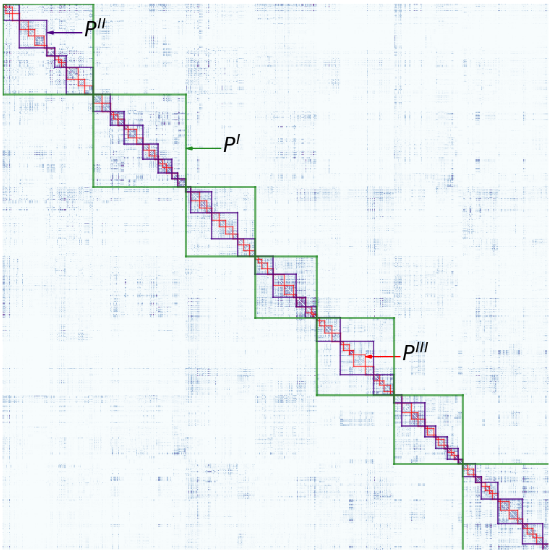}}
\label{pic:hierarchy}
\caption{Adjacency matrix of a sample continuous connectome. Rows and columns are reordered according to partition of the third hierarchical level. Boxes of different color represents clusters of different hierarchical levels. $P^{\text{I}}$ clusters are obtained first, next we reapply clustering on each detected $P^{\text{I}}$ cluster and obtain $P^{\text{II}}$. This is repeated once more to obtain $P^{\text{III}}$}
\end{figure}
We follow the hierarchical brain concept \cite{modular_brain}, repeating the clustering procedure iteratively. After the initial parcellation, we further cluster each individual parcel as an independent graph. In this work, we repeat the process three times. For each ($i$'th) continuous connectome this procedure yields a three-level hierarchically embedded partition: $P^{\text{I}}_i, P^{\text{II}}_i, P^{\text{III}}_i$, (see \textbf{Figure 1}).

\subsection{Consensus clustering} In order to obtain a unified parcellation for all subjects, we use consensus clustering. The concept was developed for aggregating multiple partitions of the same data into a single partition. We define the \textit{average} partition over all individual partitions $\{P_i\}$ as: 

\begin{equation}
\label{eq:consensus_loss}
\bar{P} == \text{argmin}_P \sum_i d(P, P_i),
\end{equation}
 where $\bar{P}$ is used to denoted desirable \textit{average} partition, $K$ is a number of averaged partitions, $d(P_i, P_j)$ is a distance measure between two partitions and we want to minimize average distance from $\bar{P}$ to all given partitions $P_i$. All partitions are represented by a vector of length $M$, where $M$ is a number of clustered objects (vertices of a graph in our case). It contains values from $1$ up to $N$, where $N$ is a number of clusters (parcels).
 This task is generally NP complete \cite{cl_ensemble}, but there are many approximate algorithms. We use two approaches: Cluster-based Similarity Partitioning Algorithm (cspa) \cite{cl_ensemble_cspa} and greedy algorithm from \cite{cl_ensemble_HE}. 

CSPA defines a similarity between data points based on co-occurrence in a same cluster across different partitions, and then partitions a graph induced by this similarity. Specifically, given multiple partitions $P_1,\ldots P_K$ of a data points $x_1,\ldots x_M$. One can define similarity between points $x_i, x_j$ as follow:
\begin{equation}
S(x_i, x_j) = \sum_{k=1}^{K} \delta(P_k(x_i), P_k(x_j)),
\end{equation}
Here $\delta$ is Kroneker delta. Thus $S(x_i, x_j)$ is just number of partitions in which points $x_i$ and $x_j$ were in the same cluster. Next we build a graph, with nodes correspond to data points and edge between node $x_i$ and $x_j$ is equal to $S(x_i, x_j)$. We the partition this graph into communities using some clustering algorithm and the resulting partition is our clustering consensus partition.

Another way to find such \textit{average} clustering is to optimize loss function given by Equation \ref{eq:consensus_loss}.

The authors of \cite{cl_ensemble_HE} propose a greedy approach (Hard Ensemble - HE). Given multiple partitions $P_1 \ldots P_K$ it combines them iteratively, first it finds average of $\bar{P}_{1,2} = \min_{\bar{P}} (d(\bar{P},P_1) + d(\bar{P},P_2))$, next average of $P_{1,2}$ and $P_3$  and so on. As a measure of distance the authors take the average square distance between membership functions:

\begin{equation}
d(P_i, P_j) = \frac{1}{N} \sum_{k=1\ldots N} ||p^k_i - p^k_j||^2,
\label{eq:dist}
\end{equation}
Exclusively for this definition we use another way to encode object's memberships: $P_i$ is a matrix of size $M \times N$ (number of objects times number of clusters)
\begin{equation}
P_i^{m,n} = 
\begin{cases} 
	1 & \mbox{if } m \mbox{'th object belongs to } n \mbox{'th cluster}\\
	0 & \mbox{otherwise.}
\end{cases} 
\end{equation}
In Equation \ref{eq:dist} $p_i^k$ and $p_j^k$ are $k^{\text{th}}$ rows of memberships matrices $P_i$ and $P_j$ respectively. They correspond to membership
vector of the $k^{\text{th}}$ object. Since we are looking for disjoint clusters, only a single element of such row vector is equal to $1$. This representation is defined up to any column permutation $\pi$ of matrix $P$, thus the optimization procedure is done subject to all possible column permutations.




\subsection{Comparison metrics} Once we find individual partitions and combine them into an average partition, we want to access their quality. We use two different approaches.

First, we compare representation strength of different parcellations by measuring distance between original $\lambda(x,y)$ and its piece-wise approximation given by:  
\begin{equation}
\gamma(x,y) = \frac{1}{|E_i||E_j|} \mathcal{C}(E_i, E_j),
\end{equation}
where $x\in E_i$ and $y\in E_j$.
Natural way to compare two statistical distributions is to measure distance between their probability density functions, we will use 
Kullback-Leibler divergence \cite{KL_div}. For two probability distributions with densities $\lambda(x)$ and $\gamma(x)$  the KL divergence is:
\begin{equation}
KL(\lambda,\gamma)=\int_{-\infty}^{\infty}\lambda(x)\log\frac{\lambda(x)}{\gamma(x)}dx,
\end{equation}
It takes values close to 0 if two distributions are equal almost everywhere. Similar but 
symmetrized version of KL divergence is Jensen-Shannon divergence \cite{JS}. Again for two probability distributions with densities $\lambda(x)$ and $\gamma(x)$ it is given by:
\begin{equation}
JS(\lambda,\gamma)=\frac{1}{2}(KL(\lambda,r) + KL(\gamma,r)),
\end{equation}
where $r(x)=\frac{1}{2}(\lambda(x)+\gamma(x))$. 

Second, we compare performance of different parcellations on a gender classification task. We use Logistic Regression model with (small) $l_1$ regularization on a vectors of edge weights (the upper triangle of adjacency matrix excluding diagonal). Classification perfomance is measured in terms of ROC AUC score, which is typical for binary classification tasks.

Finally, in order to quantify goodness of consensus clustering and access hemisphere symmetry we use Adjusted Mutual Information \cite{clustering_comparison}. It measures similarity between two partitions, with value 1 corresponds to identical partitions and values close to zero for partitions that are very different.

Using AMI we access ensemble goodness (how good clustering ensemble algorithm combines multiple partitions) using modified \ref{eq:consensus_loss}:
\begin{equation}
\text{Ensemble goodness} = \sum_i^{K} \text{AMI}(\bar{P}, P_i),
\end{equation}
We compute parcellation symmetry by comparing hemisphere parcels (labels):
\begin{equation}
\text{Symmetry} = \text{AMI} (\bar{P}_{\textbf{LH}}, \bar{P}_{\textbf{RH}}).
\end{equation}


\section{Experiments}
\subsection{Data description} We use construct continuous connetocmes of 400 subjects from the Human Connectome Project S900 release \cite{HCP} following \cite{ConCon}. We use an icosahedral spehrical sampling, at a resolution of $10242$ mesh vertices per hemisphere. We used Dipy's implementation of constrained spherical deconvolution (CSD) to perform probabilistic tractography. Prior to clustering, we exclude all mesh vertices that were labeled by FreeSurfer as corpus callosum or cerebellum.

\subsection{Experimental pipeline}
Our experiments are summarized as follows:

\begin{enumerate}
\item For each subject we reconstruct its Continuous Connectome. 
\item For each Continuous Connectome we iteratively run Louvain clustering algorithm, as described above. Subgraphs of having less then 1 percent of original graph vertices were not divided. 
\item Next we aggregate individual subject partitions and obtain consensus clustering.  Aggregation was done over 400 HCP subjects. Further, after finding the optimal parcellation, we obtain two parcellations based on two disjoint sets of 200 HCP subjects in order to compute reproducibility.
\item We aggregate partitions of the same level (I-II-III) using CSPA and HE.  
\item We compare obtained partitions between themselves and with FreeSurfer's Desikan-Killiani parcellation using Kullback-Leibler and Jensen-Shannon divergence. We compute goodness of an ensemble and parcellation symmetry using AMI. 
\item We compare performance of simplified connectomes on a binary classification task using Logistic Regression with $l_1$ penalty. Classification results are measured in terms of ROC AUC score, with averaging over $10$ cross-validation folds.
\end{enumerate}

\begin{table}
\label{tab:results}
\scriptsize
\centering
\begin{tabular}{C{1.8cm}C{1.3cm}C{1.3cm}C{1.5cm}C{1.3cm}C{1.3cm}C{1.3cm}C{1.3cm}}
\hline
\\
 & cspa $P^{\text{I}}$ & cspa $P^{\text{II}}$ & cspa $P^{\text{III}}$ & HE $P^{\text{I}}$ & HE $P^{\text{II}}$ & HE $P^{\text{III}}$ & DKT\\
\\
\hline
\\
KL&$1.22 \pm .07$ &$1.18 \pm .07$ & $1.15 \pm .07$ & $1.16 \pm .07$ & $.86 \pm .05$ & $\mathbf{.66 \pm .04}$& $.83 \pm .05$ \\
\\
JS& $.20 \pm .00$&$.19 \pm .00$ &$.19 \pm .00$ &$.20 \pm .00$ & $.17 \pm .00$ & $\mathbf{.14 \pm .00}$& $.16 \pm .00$ \\
\\
\hline
\hline
\\
Gender Classification& $.63\pm .04$ & $.64 \pm .04$ &$.69 \pm .03$ & $.64\pm .03$ & $.75 \pm .03$ & $\mathbf{.86 \pm .02}$ & $.81 \pm .03$ \\
\\
\hline
\hline
\\
Hemisphere symmetry& $.15$&$.24$ &$.32$ &$.26$ &$.55$ &$\mathbf{.66}$& $.64$   \\
\\
Ensemble goodness& $.47\pm .06$&$.40\pm.02$ &$.35\pm .00$ &$.53\pm.05$ &$.64\pm .02$ &$\mathbf{.70\pm.01}$ & $-$  \\
\\
\hline
\\
Number of ROIs &5&7&8&7&30&83&68
\\
\hline
\\
\end{tabular}
\caption{All results are rounded to 2 significant digits. Where it possible results are reported with standard deviation. Best result in each row is colored. KL, JS divergences, \textbf{lower is better}; binary Gender Classification was measured in terms of ROC AUC score, higher - better;
Ensemble goodness and Hemisphere symmetry were measured using AMI, Ensemble goodness is an average AMI between consensus partition and all individual partitions, higher - better.}
\end{table}

\begin{figure}[!h]
\center{\includegraphics[width=1\linewidth]{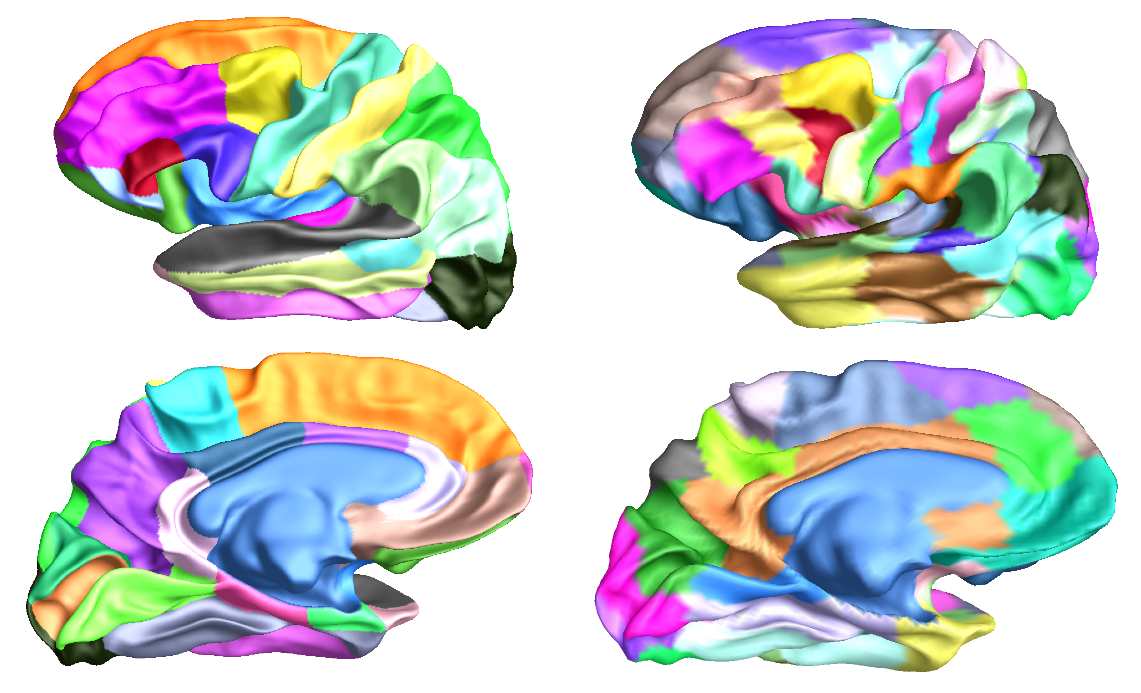}}
\label{pic:hierarchy}
\caption{Left column: Desikan-Killiany parcellation. Right column: HE $P^{\text{III}}$ parcellation. Lateral and Medial views, left hemisphere.}
\end{figure}

\subsection{Results}
Table 1
represent all comparison results. First we can see that CSPA algorithm failed to find good clustering ensemble which result in poor classification performance and high KL and JS divergences. Greedy algorithm performed on $P^{\text{III}}$ on the other hand outperforms standard Desikan atlas accross all comparison metrics (except number of parcels, 68 versus 83). Surprisingly, greedy ensemble of second level partition ($P^{\text{II}}$) performs comparatively with Desikan , despite having twice as lower number of parcels (30 versus 68).

Another interesting property that we get automatically is parcellation symmetry. Our clustering algorithm known nothing about brain topology (all information was contained in graph connectivity), still reconstruct parcellations which are highly symmetrical. For standard Desikan atlas hemisphere symmetry is $0.64$, and for our best parcellation this value even higher ($0.66$), and still remains quite high for second level partition ($0.55$).

Finally we check if our best ensemble parcellation, which combines $400$ individual partitions is stable. We split $400$ subjects into $2$ groups of $200$ subjects and independently combine their partitions. We compare resulting parcellations: $\bar{P}_{1,200}$ and $\bar{P}_{201,400}$  between themselves and with original $\bar{P}$ (which is an ensemble of all $400$ subjects) again using Adjusted Mutual Information. Both $\bar{P}_{1,200}$ and $\bar{P}_{201,400}$ shows AMI value greater than $0.80$ ($0.83$ and  $0.82$ respectively) when compare with $\bar{P}$, they also highly similar between themselves.

\section{Conclusion}
We have presented an approach for generating unified connectivity-based human brain atlases bases on consensus clustering. The method is based on finding a pseudo average over the set of individual partitions. Our approach outperforms standard a anatomical parcellation on several important metrics, including agreement with dense connectomes, improved relevance to biological data, and even improved symmetry. Because our approach is entirely data driven an requires no agreement between individual parcellation labels, it combines both the flexibility of individual parcellations and the interpretability of simple unified atlases.

\end{document}